# Artificially engineered nanostrain in FeSe$_x$Te$_{1-x}$ superconductor thin film for enhancing supercurrent

*Sehun Seo, Heesung Noh, Ning Li, Jianyi Jiang, Chiara Tarantini, Ruochen Shi, Soon-Gil Jung, Myeong Jun Oh, Mengchao Liu, Jongmin Lee, Genda Gu, Youn Jung Jo, Tuson Park, Eric E. Hellstrom, Peng Gao, & Sanghan Lee\**

*Corresponding author; e-mail: sanghan@gist.ac.kr

## Abstract

Although nanoscale deformation, such as nanostrain in iron chalcogenide (FeSe$_x$Te$_{1-x}$, FST) thin films, has attracted attention owing to the enhancement of general superconducting properties, including critical current density ($J_c$) and critical transition temperature, its formation has proven to be an extremely challenging and complex process thus far. Herein, we successfully fabricated an epitaxial FST thin film with uniformly distributed nanostrain by injection of a trace amount of CeO$_2$ inside FST matrix using sequential pulsed laser deposition. Using transmission electron microscopy and geometrical phase analysis, we verified that a trace amount of CeO$_2$ injection forms nanoscale fine defects with a nanostrained region, which has a tensile strain ($\varepsilon_{zz} \cong 0.02$) along the *c*-axis of the FST matrix. The nanostrained FST thin film achieves a remarkable $J_c$ of 3.5 MA/cm$^2$ for a self-field at 6 K and a highly enhanced $J_c$ under the entire magnetic field with respect to a pristine FST thin film.

# Introduction

Superconductors are essential materials for high magnetic field applications such as in nuclear fusion energy devices, magnetic resonance imaging (MRI), and superconducting magnetic energy storage systems. In recent years, iron-based superconductors (FeSCs) have attracted attention for high magnetic field applications because of their high upper critical field ($H_{c2}$) and low magnetic anisotropy ($\gamma$).[1,2] Moreover, the FeSCs epitaxial thin films have demonstrated enhanced overall superconducting properties compared with corresponding bulk materials.[3-9] Among several FeSCs, iron chalcogenides, (FeSe$_x$Te$_{1-x}$, FST) which possess a simple PbO type and layered-like structures[10], are excellent candidates for use as a practical superconducting material for several reasons. First, the critical transition temperature ($T_c$) of FST abruptly increases due to an increasing Se ratio with the suppression of phase separation, which is generally observed in Se-rich FST bulk, when FST is fabricated as an epitaxial thin film.[11,12] In addition, the FeSe monolayer can achieve a $T_c$ of 100 K, which is the maximum $T_c$ for FeSCs.[13] Second, FST thin films exhibit a promising critical current density ($J_c$) greater than 1 MA/cm$^2$, at the self-field regardless of the substrates including a coated conductor substrate.[5,14] This indicates that these films can potentially be used as a superconductor tape. However, $J_c$ enhancement via an artificial pinning center is a critical requirement for FST to be used in high magnetic field applications.

Several approaches have been used to improve $J_c$ in FST to date such as the use of a buffer layer,[14] oxygen annealing,[15] and ion irradiation.[16,17] In particular, low energy proton irradiation (190 KeV) is an effective method for this purpose because it causes a nanoscale cascade defect accompanied by nanostrain which enhances both $T_c$ and $J_c$ simultaneously in an FST thin film.[16] However, proton irradiation is a complicated *ex-situ* process which is not suitable for practical application. Therefore, a straightforward *in-situ* process for forming artificially controlled nanostrains is necessary to improve $J_c$ in FSTs.

Nanostrain has been generated via the introduction of various defect formations to date. For example, the insertion of a desired material with a slightly different lattice constant can induce strain through the formation of a secondary phase;[18,19] further, the doping of certain elements can generate lattice change with nanoscale strain.[20] The formation of a cascade[16] or point defect[21] by ion irradiation also induces deformation of a lattice with a nanoscale defect. In FST thin films, provided that not only large-scale and excessive amounts of defects can degrade the entire superconducting matrix in FST thin films,[22] but also FST has a short coherence length (~2 nm),[23] the formation of minimal size defects for inducing nanostrain is required to prevent Cooper pair breaking while improving $J_c$.

Herein, we report that nanostrain was successfully formed in an epitaxial FST thin film through the formation of the minimal nanoscale defects in FST thin film using sequential pulsed laser deposition (S-PLD), which can insert artificially the desired material while fabricating an epitaxial thin film.[4,24,25] We injected the precisely controlled trace amounts of $CeO_2$ to minimize the residual insertion materials to form nanoscale defects. The reason for using $CeO_2$ as an insertion material is that $CeO_2$ exhibits not only good chemical stability but also the compatible in-plane lattice constant with FST,[14,26-28] and hence the degradation of superconductivity can be minimized rather than insertion of other oxide even if the residual $CeO_2$ exists in FST. The crystallinity and structure for the $CeO_2$ injected FST (Ce-FST) was confirmed using X-ray diffraction (XRD) measurement. The nanostrain was verified using an atomic resolved scanning transmission electron microscopy (STEM) with geometrical phase analysis (GPA). The nanostrained Ce-FST thin film exhibits a significantly enhanced $J_\text{c}$ compared with that of a pristine FST (P-FST) thin film.

## Materials and Methods

### Sample preparation

We fabricated the both P-FST and Ce-FST thin films on a (001)-oriented $CaF_2$ substrate by PLD using a KrF (248 nm) excimer laser (Coherent, COMPEX PRO 205F) in a vacuum of $2\times10^{-5}$ Pa at 400 °C. We used a $FeSe_{0.45}Te_{0.55}$ target, which was made by an induction melting method. FST thin films were grown using a laser energy density of 3 J/cm$^2$, a repetition rate of 3 Hz, and target to substrate distance of 4 cm.

The method for fabrication of Ce-FST thin film is as follow. We first deposited 20 nm (445 laser pulses) FST layer on $CaF_2$ substrate. Then, $CeO_2$ was deposited on FST layer. These processes were repeated four times in total, and finally, an FST layer is deposited on top surface. The total thickness of all FST thin films was 100 nm (2225 laser pulses). $CeO_2$ was deposited between FST layers with a dependence on $p$ (2, 5, 10, and 20); $p$ is the number of the laser pulses of the inserted $CeO_2$ (laser energy density of 1.5 J/cm$^2$ and a repetition rate of 1 Hz). The target changing time between the FST and $CeO_2$ targets was 10 seconds which is the drive time when the laser is turned off and then on again. The composition of FST thin film is considered as $FeSe_{0.7}Te_{0.3}$, approximately, based on our previous report.[11]

### Characterization

To characterize the crystal structure, the $\theta$-$2\theta$ scan, azimuthal phi scan, and rocking curve were measured using a four-circle XRD (PANalytical, X'Pert pro, $\lambda$ = 1.5406 Å).

We also performed an additional $\theta$-$2\theta$ scan at the beamline 3A in the Pohang Accelerator Laboratory with a six circle XRD ($\lambda$ = 1.148 Å). STEM images and EDS were obtained in a $C_s$-corrected FEI Titan Themis G2 at an accelerating voltage of 300 KV with a beam current of 70 pA, a convergence semi-angle of 15 mrad, and a collection semi-angle snap in the range of 80-379 mrad. To obtain GPA maps for Ce-FST and P-FST thin films, same parameters were used for all calculation (same Fourier vector, resolution: 5 nm, smoothing: 10 nm, color scale: -0.1~0.12). The resistivity-temperature measurement was performed using a physical property measurement system (PPMS, Quantum Design). $T_c^{onset}$ and $T_c^{zero}$ were determined using the 0.9 $\rho_n$ criterion and 0.01 $\rho_n$ criterion, respectively, where $\rho_n$ is resistivity at 23 K. Magnetization $J_c$ was measured using a vibrating sample magnetometer (VSM, Oxford) by applying a magnetic field perpendicular to the film. It was estimated using a Bean model for thin film: $J_c = 15\Delta M/Vr$, where V is the thin film volume in cubic centimeter, r is the equivalent radius of the sample size ($\pi r^2 = a \times b$; a and b are width and length of the sample), and $\Delta M$ is the width of magnetic moment from the M-H loop (for further information, see Supplementary Fig. S1). Transport $J_c$ is obtained by direct transport measurement of the patterned FST sample using a standard four-probe method.

## Result and discussion

### Crystalline phases

Figure 1 shows the schematic diagram of two different Ce-FST thin films. First, if the amount of inserted $CeO_2$ is very small (2 *p*, below 0.5-unit cell), nanoscale defects can be formed inside the FST thin film, not the $CeO_2$ layer, given that the inserted 2 *p* $CeO_2$ is the infinitesimal amount which is insufficient to form nucleation clusters or layers inside an FST. These nanoscale defects can generate the nanostrain inside FST thin film (left side in Fig. 1). The mechanism will be discussed later in detail. If the inserted $CeO_2$ (20 *p*, 2.5 unit cell) is sufficient to form a $CeO_2$ layer in an FST thin film, the $CeO_2$ layer is formed between the respective FST layers without nanostrain (right side in Fig. 1). Thus, 20 *p* Ce-FST thin film forms superlattice FST thin film with $CeO_2$.

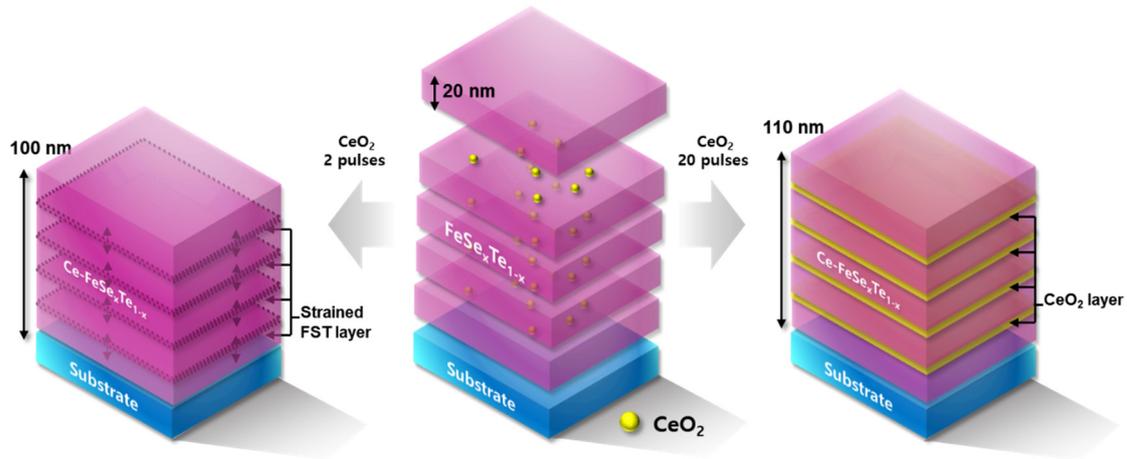

**Figure 1**. Schematic of the two types of Ce-FST thin films. The infinitesimal $CeO_2$ (2 *p*) is injected at the interface between each FST layer to form nanostrained FST (left side). The sufficient $CeO_2$ (20 *p*) is injected at the interface between each FST layer to form $CeO_2$ layer (right side). The thickness of each FST layer is 20 nm, and the total thickness was 100 to 110 nm.

We performed a $\theta$-$2\theta$ scan using XRD to identify the out-of-plane crystalline quality of the Ce-FST thin films. Figure 2a shows the out-of-plane $\theta$-$2\theta$ XRD patterns of the P-FST and Ce-FST thin films dependent on *p* (2, 5, 10, and 20). The $\theta$-$2\theta$ scans clearly show only (00*l*) peaks for Ce-FST thin films with $CaF_2$ (00*l*) peaks. However, there are no other phase peaks despite the periodically injected $CeO_2$ because the amount of the inserted $CeO_2$ is significantly small to be measured in XRD. Figure 2b shows the enlarged section of Fig. 2a close to the (001) peak of the Ce-FST thin films. The (001) peak of 2 *p* Ce-FST is noticeably shifted more to the left than that of P-FST, indicating that 2 *p* Ce-FST receives the tensile strain along the *c*-axis. Intriguingly, the degree of shift of the (001) peak returns to zero with increasing the *p*. This indicates that the strain is relaxed in Ce-FST thin films with an increase in *p*. Also, the same shift tendency is observed in other 00*l* peaks in Ce-FST thin films (for further information, see Supplementary Fig. S2).

We additionally measured the *θ-2θ* scan for the 2 *p* Ce-FST and 20 *p* Ce-FST thin films using a synchrotron based XRD to further verify for the crystalline structure (for further information, see Supplementary Fig. S3). As shown in Fig. S3, the only (00*l*) peaks of both 2 *p* Ce-FST and 20 *p* Ce-FST thin films are observed, and the (001) peaks display satellite peaks which have been generally observed in superlattice thin films.[4] Since 20 *p* Ce-FST thin films can have superlattice structure with $CeO_2$, satellite peaks can be observed. However, in 2 *p* Ce-FST thin film, it is difficult to form superlattice structure with forming intact $CeO_2$ layer in FST matrix because a trace amount of $CeO_2$ was injected in the FST matrix. Thus, we speculated that the satellite peaks of 2 *p* Ce-FST thin film are due to small changes such as the nanostrain at the interface between the FST layers.

Additionally, we measured the rocking curve for the (001) reflection of both P-FST and 2 *p* Ce-FST thin films using four circles XRD to compare the out-of-plane crystalline quality and the mosaicity (Figure 2c). The calculated full width at half maximum (FWHM) of the (001) reflections was 0.67° for the 2 *p* Ce-FST and 0.55° for the P-FST, respectively. The difference in the FWHM between P-FST and 2 *p* Ce-FST thin films is minimal, and the FWHM of 2 *p* Ce-FST is similar to other reported $FeSe_xTe_{1-x}$ thin films.[5,9,11] This indicates that the 2 *p* Ce-FST thin film was well grown along the *c*-axis despite the insertion of the oxide materials into the FST matrix.

To confirm in-plane texture and epitaxial quality, we performed an azimuthal phi scan using four circles XRD. Figure 2d indicates the azimuthal $\phi$ scan of the (113) peak from the CaF$_2$ substrate and the (112) peak from the 2 *p* Ce-FST thin film. The $\phi$ scan of the 2 *p* Ce-FST thin film shows clear four peaks which have 90° intervals without extra broader intermediate-angle peaks. These peaks are at 45° with respect to the CaF$_2$ (113) peaks because the [100] FST is parallel to the [110] CaF$_2$. These results indicate that 2 *p* Ce-FST has the characteristic of a genuine epitaxial film with excellent in-plane texture.

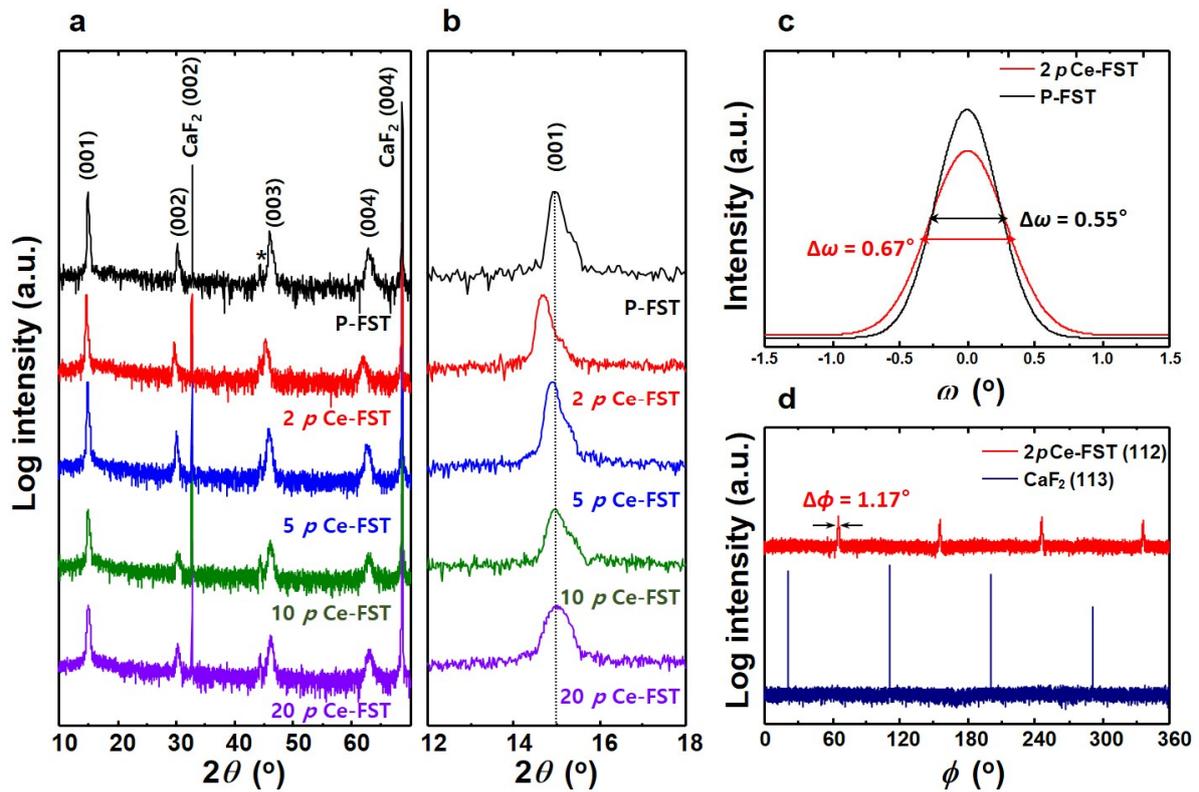

**Figure 2.** (a) Out-of-plane $\theta$-$2\theta$ XRD pattern of the P-FST and Ce-FST thin films grown on CaF$_2$. (b) Enlarged FST (001) reflection on Fig. 2a. (c) Rocking curve of (001) reflection of the 2 *p* Ce-FST and the P-FST thin films. Here, $\Delta\omega$ indicates the FWHM of the (001) reflection of both films. (d) Azimuthal $\phi$ scan of the 2 *p* Ce-FST (113) and CaF$_2$ (112). * peak indicates an artefact peak from XRD equipment.

**Strain analysis**

To determine the nanoscale strain caused by the infinitesimal CeO$_2$ injection at the interface of each FST layers, we analyzed atomically resolved-STEM images for the 2 *p* Ce-FST thin film. Figure 3a shows a cross-sectional atomically resolved-STEM image of the 2 *p* Ce-FST thin film. As shown in Fig. 3a, no other dominant phases such as CeO$_2$ particle are observed except for the FST phase. Although we double check to find CeO$_2$ particles in 2 *p* Ce-FST thin film, there are no CeO$_2$ layers, CeO$_2$ particles, or large scale defects (for further information, see Supplementary Fig. S4). Whereas, fine bright lines are observed with vertically 20 nm intervals in 2 *p* Ce-FST thin film. To analyze the fine bright lines in 2 *p* Ce-FST thin films, we performed GPA based on the atomically resolved-STEM image of Fig. 3a. GPA is generally used to show strain distribution and to determine deformation of the lattice constant in crystalline structure.[29] Figure 3b shows an extracted strain map of out-of-plane strain ($\varepsilon_{zz}$) for the identical region in Fig. 3a. The GPA map undoubtedly displays the strained region with vertically 20 nm intervals; The thickness of the nanostrained region is 5 ~ 10 nm, approximately. To further analyze the strain, we plotted the line profile of strain for 2 *p* Ce-FST thin film based on GPA result. As shown in Fig. 3c, nanostrains are observed with vertically 20 nm intervals. This nanostrain is a tensile strain ($\varepsilon_{zz} \cong 0.02$) along the *c*-axis, and this nanostrain position is in good agreement with the expected site where we intentionally inserted CeO$_2$.

For a more accurate comparison, we performed STEM analysis of the P-FST and 20 *p* Ce-FST thin films. The P-FST thin film exhibits a relatively clear phase as shown in Fig. 3d; there is no particular strain field in the out-of-plane GPA strain map from the atomically resolved-STEM image of the P-FST (Figure 3e). Figure 3f shows the line profile of out-of-plane strain for P-FST thin film. As shown in Fig. 3f, the strain in P-FST thin film fluctuates around zero.

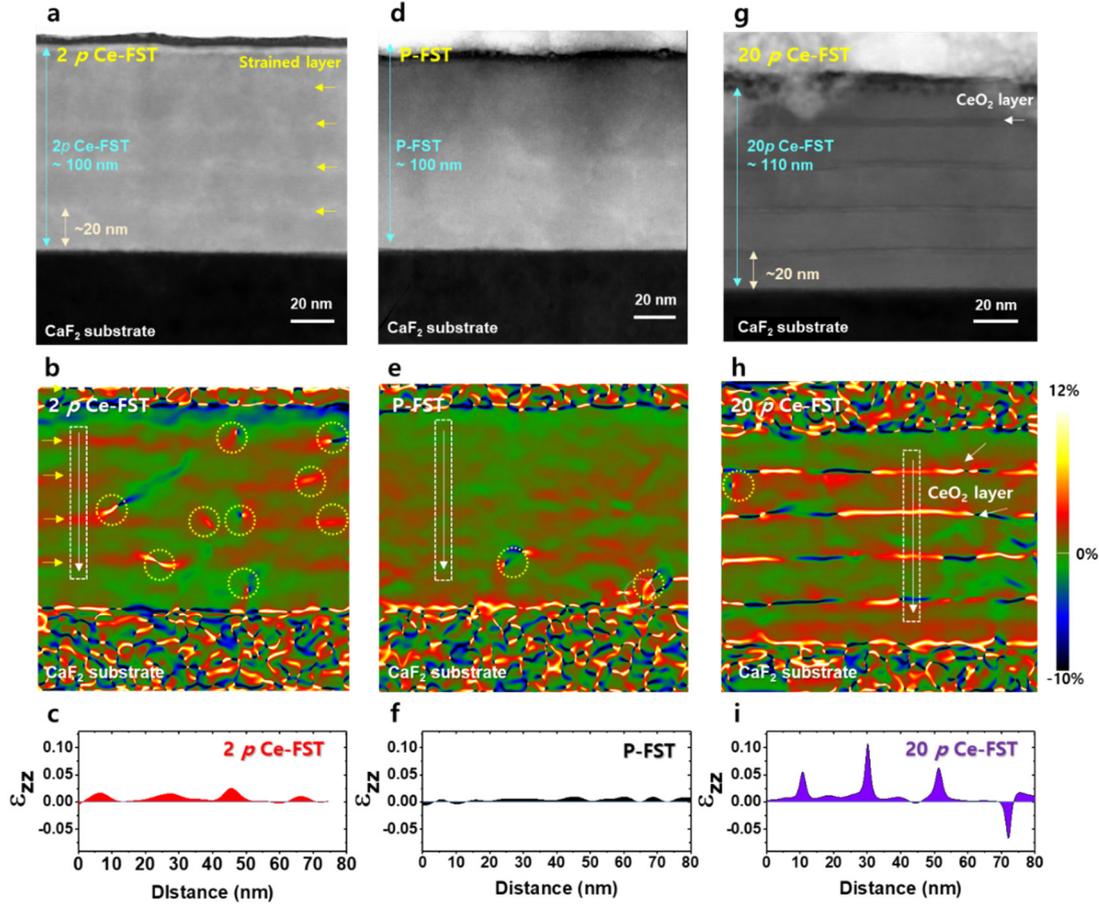

**Figure 3**. (a) Cross-sectional HAADF-STEM image of the 2 *p* Ce-FST thin film. (b) Map of out-of-plane strain for the 100 nm 2 *p* Ce-FST thin film determined by GPA of STEM images from the same area on Fig. 3a. Yellow arrows indicate nanostrained FST region. (c) The line profile of strain in the white rectangular region of Fig. 3b. (d) Cross-sectional STEM image of the P-FST thin film. (e) Map of our-of-plane strain for the 100 nm P-FST thin film determined from GPA of STEM images from the same area on Fig. 3d. (f) The line profile of strain in the white rectangular region of Fig. 3e. (g) Cross-sectional STEM image of the 20 *p* Ce-FST thin film. (h) Map of our-of-plane strain for the 20 *p* Ce-FST thin film determined from GPA of STEM images from the same area on Fig. 3g. (i) The line profile of strain in the white rectangular region of Fig. 3h. In GPA images, dashed circles represent lattice distortion points. Strain scale bar of GPA is the same in all GPA images. Both the strain contrast at the CeO$_2$ layer in GPA map of Fig. 3h and the strain peaks of line profile in Fig. 3i are artifact which is caused by structure difference between FST (PbO type structure) and CeO$_2$ (fluorite structure).

The 20 *p* Ce-FST thin film exhibits a clear $CeO_2$ layer between the 20 nm interval of FST layers (Figure 3g). Figure 3h shows the out-of-plane GPA image of the 20 *p* Ce-FST thin film. Figure 3i shows the line profile of out-of-plane strain for 20 *p* Ce-FST thin film. The large strain contrast at $CeO_2$ layer in Fig. 3h and 3i is artifact which is caused by structure difference between FST and $CeO_2$. The relatively small strain fields (<0.02) are irregularly observed near the $CeO_2$ layer in GPA map for 20 *p* Ce-FST thin film. One interesting fact is that nanostrain is observed although there is no $CeO_2$ layer or particle in the STEM image of 2 *p* Ce-FST thin film. Thus, it is important to demonstrate why the injected trace amount of $CeO_2$ forms nanostrain in the FST matrix and why there are no $CeO_2$ particles in 2 *p* $CeO_2$ FST thin film.

In general, nanostrain is induced at various types of defect perimeters.[16,18-21] Interestingly, lattice distortion points (dashed circle in GPA maps of Fig. 3) such as dislocation core and damaged FST layer are prominently observed at the nanostrain region in the GPA image for the 2 *p* Ce-FST thin film. Also, there are a few nanoscale defects which are formed irrespective of $CeO_2$ insertion in the FST thin films as shown in Fig 3d. These nanoscale defects can cause the nanostrain in FST thin film. However, it is difficult to form the nanostrain over the broad region only by these nanoscale defects because these defects form the localized strain field.[18]

To further understand the origin of the nanostrained region, we analyzed the enlarged STEM image of the nanostrained region where there are no lattice distortion points using energy dispersive spectroscopy (EDS) mapping. Figure 4a, 4b, and 4c show the different STEM images for the 2 *p* Ce-FST thin film with EDS mapping results for the respective atoms, i.e. Fe, Se, Te, Ce, and O. Fig. 4d shows the STEM image for the 20 *p* Ce-FST thin film with EDS mapping results for the respective atoms, i.e. Fe, Se, Te, Ce, and O. As shown in Fig. 4d, the $CeO_2$ layer is certainly observed in 20 *p* Ce-FST thin film. The inserted CeO₂ was epitaxially grown with a relation (001)[110]FST || (001)[100]CeO₂. Both Ce and O are strongly detected around the CeO₂ layer in EDS map for 20 *p* Ce-FST, whereas there is no Ce and O signal around nanostrain region in EDS maps for the 2 *p* Ce-FST thin film (Fig. 4a, 4b, and 4c).

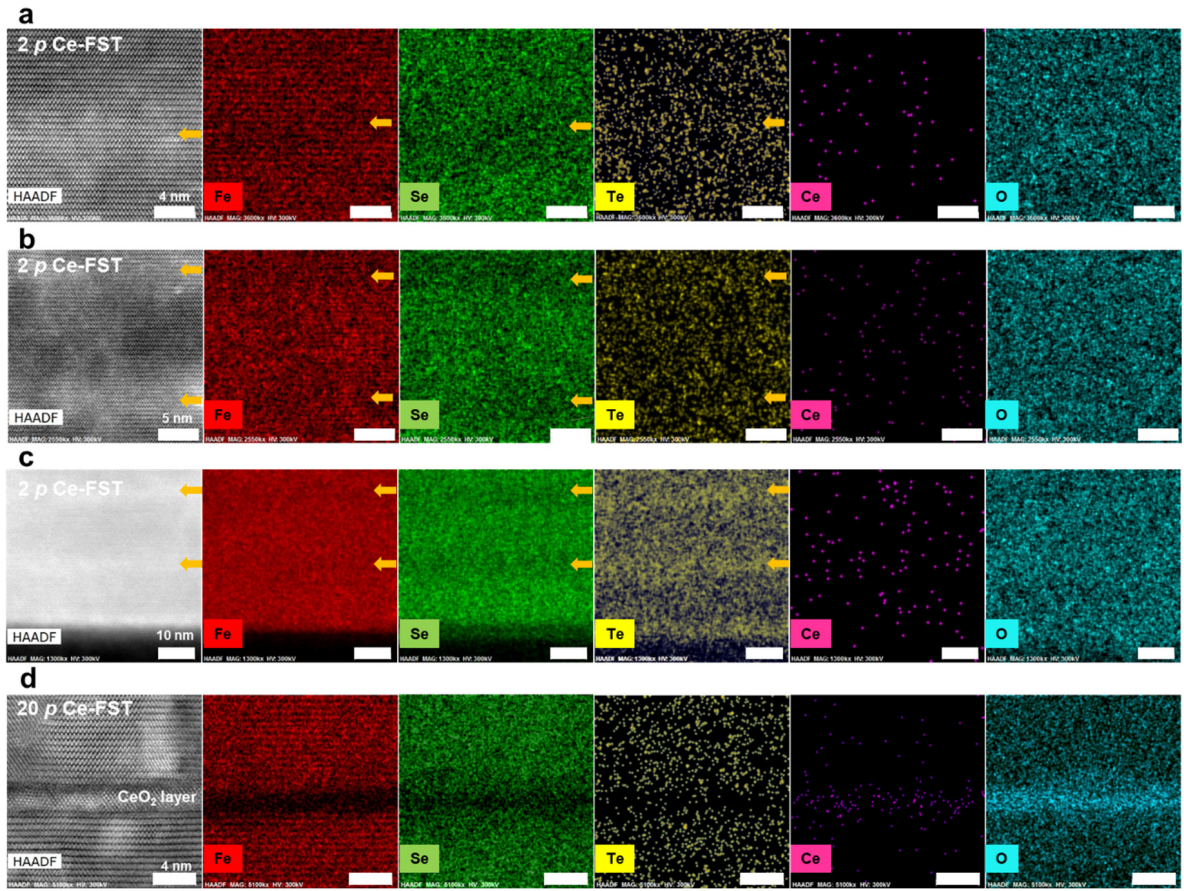

**Figure 4.** (a, b, c) Cross-sectional HAADF-STEM image of each different nanostrained region in 2 *p* Ce-FST thin film with EDS Map of the respective atoms (Fe, Se, Te, Ce, and O) from the same area of STEM images. (d) Cross-sectional HAADF-STEM image of the CeO₂ layer in 20 *p* Ce-FST thin film with EDS Map of the respective atoms from the same area of STEM image. Yellow arrows indicate the nanostrained regions.

One interesting discovery is that not only fine decrease of both Se and Fe ratio but also fine increase of Te ratio are observed at nanostrained region in the EDS maps for 2 *p* Ce-FST thin film when we analyzed these EDS maps using plot profile (for further information, see Supplementary Fig. S5). The decrease in the Se ratio can cause an increase in the lattice constant.[11] This indicates that nanostrain can be induced near the Se deficient region which is generated by infinitesimal $CeO_2$ insertion.

Thus, it is important to demonstrate why Se deficiency is observed in nanostrained region without residual $CeO_2$ particle. In the PLD system, the laser ablation of the target forms a plume which contains the ionized species with high kinetic energy. These ionized energetic species cause re-sputtering and the formation of fine defects on the surface in the initial stage before these species form the cluster or layer.[30] In this re-sputtering stage, it is impossible for the inserted $CeO_2$ to form an intact $CeO_2$ layer; instead, re-sputtering forms nanoscale defects and damaged FST layers (or transition layer). These phenomena were observed in not only our 20 $p$ Ce-FST thin film as shown in Fig. 4d, but also in other papers when the $CeO_2$ layer was deposited into or onto the FST thin film.[26,27,31] Further, Se deficiency can be generated at the re-sputtering stage, provided that the atomic ratio of FST abnormally changed during thin film growth because of instability of Fe-Se bonding.[11] Collectively, the nanostrain can be induced by nanoscale defects, such as lattice distortion points, a damaged FST layer, and Se deficiency which are formed by inserting infinitesimal amount of $CeO_2$.

Furthermore, we examined whether the formation of the nanostrain is affected by pausing time (10 s) during target exchange between the FST and $CeO_2$ targets, because Se and Te are sensitive and volatile in FST thin films.[11] The paused FST thin film was fabricated following the same fabrication process for the 2 $p$ Ce-FST thin film except for $CeO_2$ injection; the $CeO_2$ plume was screened during the laser ablation of the $CeO_2$ target. When we performed the $\theta$-$2\theta$ scan for paused FST thin film using Synchrotron-based XRD, the results indicated well-oriented (001) peaks without satellite peak, indicating that the pausing time had a negligible effect on the formation of the nanostrain in the FST matrix (for further information, see Supplementary Fig. S3).

**Superconductivity measurements**

We measured the temperature dependence resistivity to obtain the $T_c$ to compare superconducting properties between the P-FST and Ce-FST thin films (Figure 5a). The measured $T_c^{onset}$s are 21.3 K, 20.4 K, 19.0 K, 16.9 K, and 16.7 K, respectively, for the P-FST, 2 $p$ Ce-FST, 5 $p$ Ce-FST, 10 $p$ Ce-FST, and 20 $p$ Ce-FST thin films. In particular, the $T_c$s of FST thin films decrease with an increase in the $p$ of CeO$_2$. The primary reason for $T_c$ degradation in the Ce-FST thin films is the degradation of crystalline quality with the increase in the amount of the inserted CeO$_2$ (for further information, see Supplementary Fig. S6). Figure 5b, 5c, and S7 (for further information, see Supplementary S7) show the resistivity as a function of temperature up to 9 T with H//c for 2 $p$ Ce-FST, P-FST, and other Ce-FST thin films, respectively. Interestingly, the suppression of $T_c$ which is dependent on the magnetic field ($\Delta T_{c,\,field}^{zero} = T_{c,\,9T}^{zero} - T_{c,\,0T}^{zero}$) for the 2 $p$ Ce-FST thin film ($\Delta T_{c,\,field}^{zero} = 2.6$ K) is lower than that of the P-FST thin film ($\Delta T_{c,\,field}^{zero} = 3.2$ K); the measured $T_{c,\,0T}^{zero}$ and $T_{c,\,9T}^{zero}$ are 19.8 K and 16.6 K, respectively, for the P-FST thin film and 18.9 K and 16.3 K, respectively, for the 2 $p$ Ce-FST thin film. This indicates that the 2 $p$ Ce-FST thin film has a lower magnetic field dependence than the P-FST thin film although $T_c$ of 2 $p$ Ce-FST is lower than that of P-FST thin film.

We estimated $H_{irr}$ and $H_{c2}$ of the Ce-FST and P-FST thin films using the 0.01 $\rho_n$ criterion and the 0.9 $\rho_n$ criterion with $\rho_n = \rho(23\,K)$ as a function of the normalized temperature ($t = T/T_c^{onset}$), to characterize the temperature dependence of the characteristic fields. (Figure 5d). The improved $H_{irr}$ of 2 $p$ Ce-FST is indicative of the beneficial effect of the periodic nanostrained region with the nanoscale defects as pinning centers under high magnetic fields. Whereas, $H_{c2}$ and $H_{irr}$ of other Ce-FST (5, 10, and 20 $p$) are degraded after the CeO$_2$ insertion, indicating that the CeO$_2$ particle or layer can degrade $H_{irr}$ and $H_{c2}$ in an FST thin film.

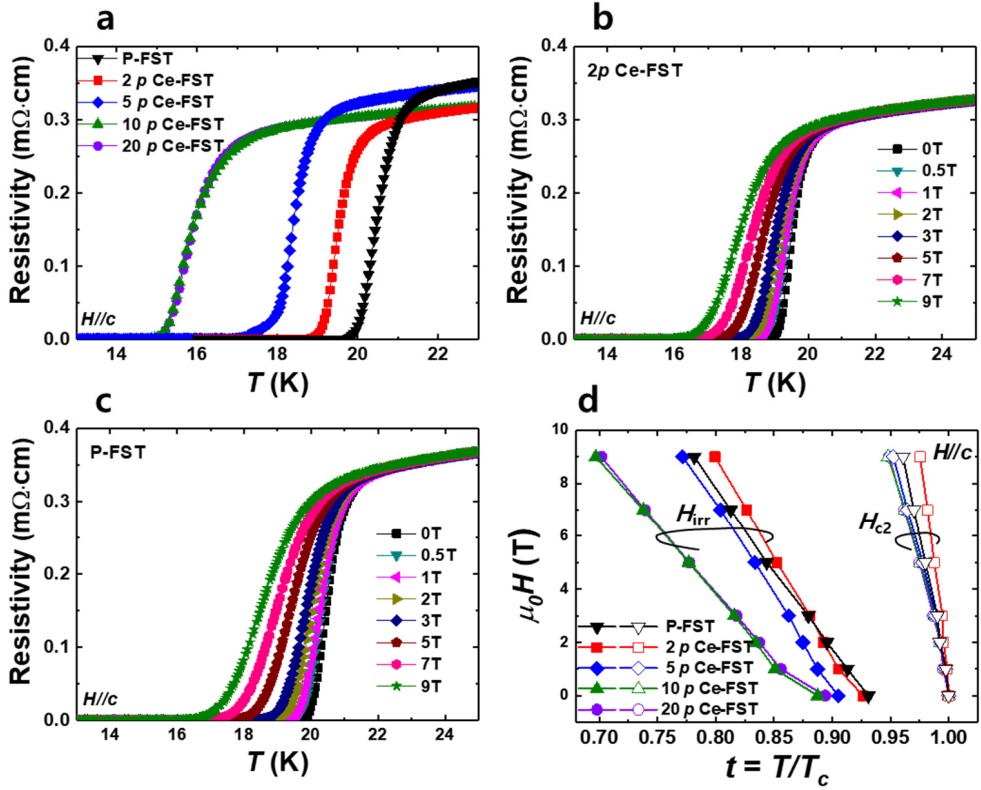

**Figure 5**. (a) Temperature dependence of the resistivity of the P-FST and Ce-FST thin films. Temperature dependence of the resistivity of (b) the 2 *p* Ce-FST thin film and (c) the P-FST thin film depending on applied magnetic fields. (d) Upper critical field [$H_{c2}$(T)] and irreversibility field [$H_{irr}$(T)] as a function of normalized temperature ($t=T/T_c$) for the Ce-FST and P-FST thin films. Each point shows the resistivity drops to 0.9 of normal resistivity ($\rho_n$) and 0.01 of $\rho_n$. $\rho_n$ is the resistivity at the normal state ($\rho$(23 K)) of P-FST and Ce-FST thin films, respectively.

The $J_c$ of both the 2 *p* Ce-FST and the P-FST thin films were measured to verify the effect of the nanostrain as a pinning center on supercurrents in the FST thin film (Figure 6). Figure 6a and 6b show the magnetic field dependence magnetization $J_c$ of the 2 *p* Ce-FST and the P-FST thin films at various temperatures (4.2 K, 7 K, 10 K and 12 K) up to 13 T (*H//c*). The magnetization $J_c$ of the 2 *p* Ce-FST thin film had a value of 3.2 MA/cm$^2$ in a self-field and 0.44 MA/cm$^2$ under 13 T, at 4.2 K. The self-field $J_c$ of the 2

*p* Ce-FST thin film is the highest value of an iron-chalcogenide superconductor to the best of our knowledge.[15, 32] The magnetization $J_c$ of the P-FST thin film had a value of 2.3 MA/cm$^2$ in a self-field and 0.23 MA/cm$^2$ under 13 T at 4.2 K. The magnetization $J_c$ of the P-FST thin film is also similar and higher when compared to other reported values.[15,16,27] The transport $J_c$ of the both P-FST and Ce-FST thin film was also measured at 6 and 10 K to verify these magnetization $J_c$ value derived using the Bean model (Figure 6c). The 2 *p* Ce-FST shows the transport $J_c$ of 3.5 MA/cm$^2$ in a self-field and of 0.44 MA/cm$^2$ under 13 T at 6 K, which is reasonably similar to the magnetization $J_c$ of the Ce-FST thin film at 4.2 K. The P-FST shows a transport $J_c$ of 0.91 MA/cm$^2$ in a self-field and of 0.10 MA/cm$^2$ under 13 T at 6 K, which is similar to the magnetization $J_c$ of the P-FST thin film at 7 K.

Additionally, $J_c$ enhancement was calculated to confirm the effect of the nanostrain in detail based on the magnetization $J_c$ (for further information, see Supplementary Fig. S8). The $J_c$ enhancement of 2 *p* Ce-FST compared with P-FST increased from 40% to 120% up to 5 T and gradually decreased for a high magnetic field. These results clearly demonstrate that the 2 *p* Ce-FST maintains a high $J_c$ under a high magnetic field as well as a low magnetic field. Furthermore, we measured the angular dependence of transport $J_c$ of the 2 *p* Ce-FST and P-FST evaluated at a constant reduced temperature $T/T_c$~0.6 to understand the pinning effects of the nanostrain with nanoscale defects. As shown in Fig. 6(d), the in-plane $J_c$ is 48% higher than the perpendicular $J_c$ in the P-FST thin films. The 2 *p* Ce-FST film shows an enhanced in-plane $J_c$ of 1.6 MA/cm$^2$ which is 60% higher than the perpendicular $J_c$ of 1.0 MA/cm$^2$ due to the in-plane pinning effect by the lateral nanostrain. Above all, the nanostrain with other nanoscale defects improves the $J_c$ of the 2 *p* Ce-FST film under all directions of magnetic field.

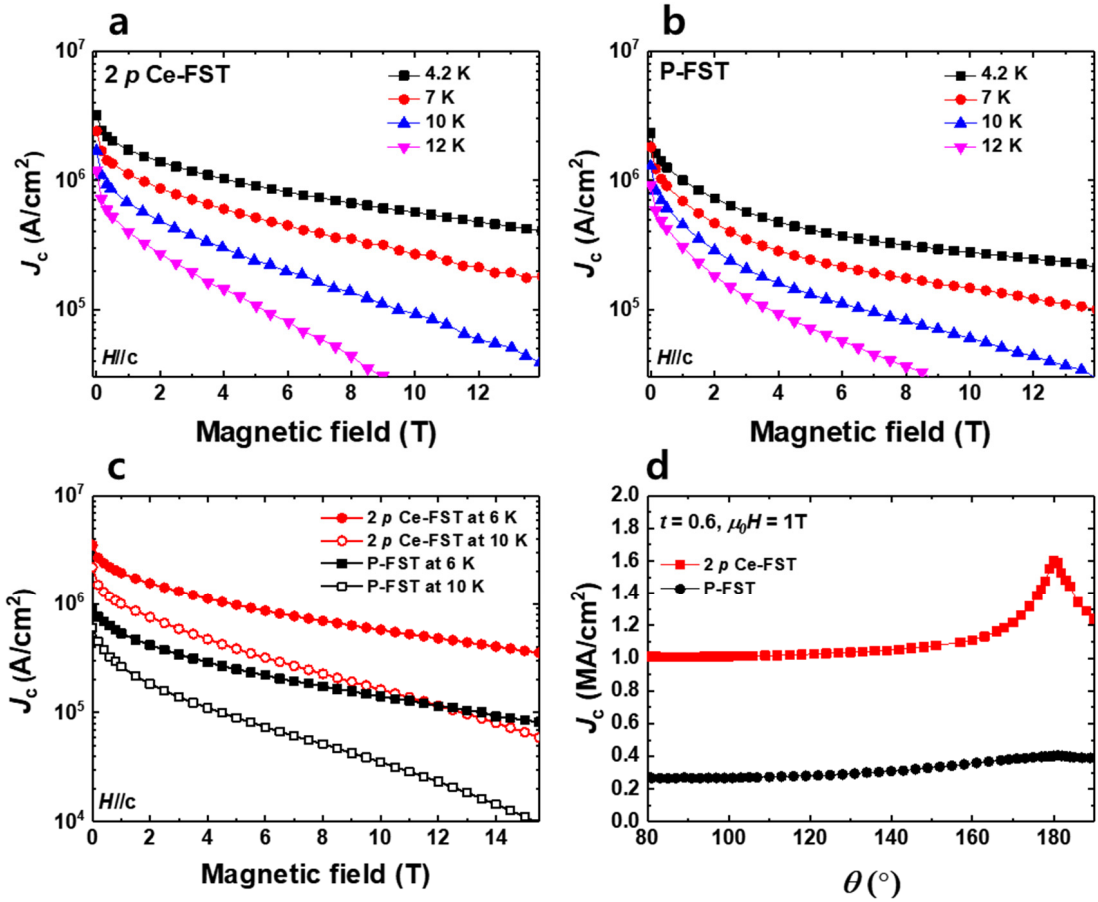

**Figure 6**. Magnetization $J_c$ as a function of the magnetic field ($H$//c) of (a) the 2 $p$ Ce-FST and (b) the P-FST thin films which were derived from half magnetization loops at different temperatures (4.2 K, 7 K, 10 K and 12 K). (c) Transport $J_c$ as a function of magnetic field ($H$//c) of both P-FST and 2 $p$ Ce-FST thin films up to 15 T at 6 K and 10 K. (d) Angular dependence of transport $J_c$ of the 2 $p$ Ce-FST and P-FST evaluated at a constant reduced temperature $t$ ($T/T_c$) ~ 0.6.

**Relationship between nanostrain and $J_c$**

Also, we plotted lattice constant, $c$, and magnetization $J_c$, at 4.2 K as a function of $p$ of the inserted $CeO_2$ to further understand the relationship between nanostrain and $J_c$ (Figure 7). The lattice constants were obtained by Nelson Riley method based on XRD results and by fast Fourier transform (FFT) analysis based on STEM images (for further information, see Supplementary S9 & S10). The magnetization $J_c$ of Ce-FST thin films (2, 5, 10, and 20 $p$) was measured at 4.2 K (for further information, see Supplementary

Fig. S11). As shown in Fig. 7, the change in $J_c$ follows the same tendency as the change of lattice constant, $c$, which represents the change in strain. This tendency demonstrates that nanostrain is responsible for the enhanced $J_c$ in the 2 *p* Ce-FST thin film.

To understand the pinning mechanism of the 2 *p* Ce-FST thin film, we plotted $J_c(t)/J_c(0)$ versus $t$ for both 2 *p* Ce-FST and P-FST thin films based on magnetization $J_c$ of both 2 *p* Ce-FST and P-FST thin films (for further information, see Supplementary S12). Both 2 *p* Ce-FST and P-FST thin films show the $\delta l$-pinning type which is caused by fluctuation of the charge-carrier mean free path. We also calculated scaled volume pinning force ($f_p$) as a function of the normalized field ($h$); $f_p$ is $F_p/F_{p\_max}$, and $h$ is $H/H_{irr}$ (for further information, see Supplementary Fig. S12). In general, the $h$ values of 0.2 and 0.33 indicate a surface pinning geometry and point pinning geometry, respectively.[33] If $J_c$ of 2 *p* Ce-FST thin film is improved by $CeO_2$ particle or other defects which cause point and volume pinning, $h$ value is shifted to 0.33. However, the $h$ of $f_p$ is approximate 0.2 in both P-FST and 2 *p* Ce-FST thin films, indicating that main pinning type in both P-FST and 2 *p* Ce-FST thin film is surface pinning.

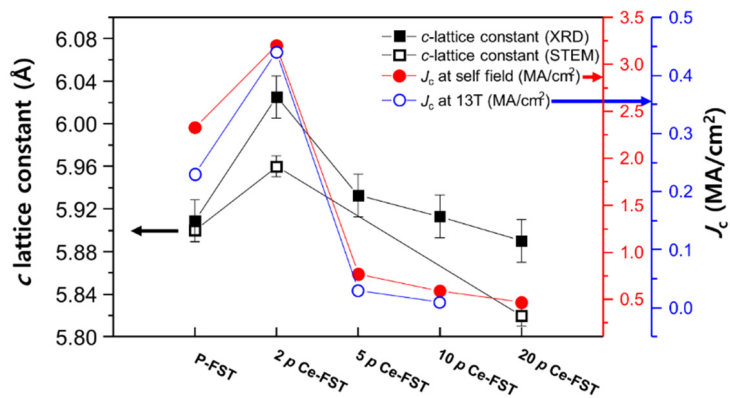

**Figure 7.** As a function of *p* in FST thin films, *c*-lattice constants which were calculated from STEM and XRD results, magnetization $J_c$ at self-field, and magnetization $J_c$ under 13 T.

In P-FST thin film which has pure FST phase without defects, the interlayer spacing between Fe-Se(Te) planes can be the intrinsic pinning center due to short coherence length of FST.[34] Since this interlayer spacing has a two-dimensional lateral gemoetry, the pinning type of interlayer spacing in P-FST thin film is surface pinning geometry. One interesting fact is that the 2 *p* Ce-FST thin film also has surface pinning geometry even though $J_c$ is relatively improved than P-FST thin film. This means that 2 *p* Ce-FST thin film has similar pinning type to compare with P-FST thin film. The difference characteristic between P-FST thin film and 2 *p* Ce-FST thin film is that the *c* lattice of 2 *p* Ce-FST thin film is expanded by nanostrain rather than P-FST thin film, indicating that interlayer spacing of 2 *p* Ce-FST thin film at nanostrained region is larger than that of P-FST thin film. Thus, the interlayer spacing of 2 *p* Ce-FST can be more effective pinning center than that of P-FST.

To evaluate the efficiency of our method, we compared the pinning characteristics with that of the previous papers.[14-16,35-38] Figure 8a shows the transport $J_c$ of 2 *p* Ce-FST and P-FST at 6 K for H//c together with the $J_c$ for other reported superconductors. The 2 *p* Ce-FST thin film exhibits a higher $J_c$ than other reported $J_c$ of FST thin films.[14-16,35-38] We also estimated the pinning force ($F_p$) to characterize the effect of nanostrain with nanoscale defects. Fig. 8b shows the magnetic field dependence of the vortex pinning force ($F_p = J_c \times B$) of both the 2 *p* Ce-FST and the P-FST thin films up to 13 T (*H*//c) at 6 K together with the reported $F_p$ of other superconductors.[14-16,35-38] The 2 *p* Ce-FST and P-FST thin films show maximum pinning force ($F_{p,max}$) of 57.8 GN/m³ under 11.5 T and 14.2 GN/m³ under 11 T at 6 K, respectively. In particular, the 2 *p* Ce-FST thin film shows ~400% higher $F_{p,max}$ than for the P-FST thin film at 13 T (*H*//c). In addition, the 2 *p* Ce-FST thin film exhibits a higher $F_p$ than that of other reported FST even though our samples were measured at relatively higher temperature of 6 K.

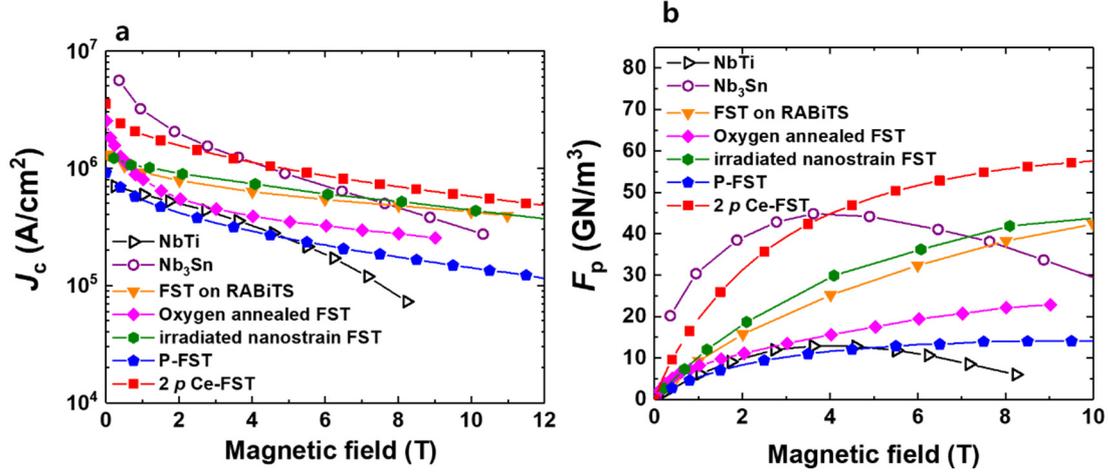

**Figure 8.** Transport $J_c$ (a) and vortex pinning force $F_p$ (b) dependence of magnetic field for several high field superconductors, including P-FST, 2 *p* Ce-FST (this work), the FST on coated conductors (RABiTS), the oxygen annealed FST, ion irradiated FST, Nb-Ti, and Nb$_3$Sn. 4.2 K $J_c$ data of high field superconductors. Except for our samples, all data are adopted from the literature.[14-16,34-38]

## Conclusions

Herein, we have successfully induced nanostrain inside FST thin films via the injection of an infinitesimal amount of CeO$_2$ using S-PLD without additional post-processing. Through STEM analysis with GPA and EDS, we demonstrate that the injected infinitesimal amount of CeO$_2$ forms nanoscale defects such as dislocation core and Se deficiency, which forms tensile nanostrain along the *c*-axis in the FST thin film. The nanostrain significantly improves the self-field transport $J_c$ of the FST thin film from 0.91 MA/cm$^2$ up to 3.5 MA/cm$^2$ at 6 K, while minimizing the degradation of the $T_c$ of the FST thin film. This study demonstrates that the formation of the nanostrain using S-PLD is significantly effective in attempting to achieve the ultimate goal of high magnetic field applications of iron chalcogenide. We also believe that this technique will be of great utility in inducing artificial nanoscale strains in other epitaxial chalcogenide thin films.

## Acknowledgements

This work was supported by the Global Research Network program (2014S1A2A2028361), the Creative Materials Discovery Program (2017M3D1A1040828), the Basic Science Research Program (2016R1D1A1B03931748), and the international cooperation program (2018K2A9A1A06069211) through the National Research Foundation of Korea (NRF) and funded by the Ministry of Science and ICT, and the Ministry of Education, and by the GRI (GIST Research Institute) project through a grant provided by GIST. P.G. is grateful for the support from the National Key Research and Development Program of China (2016YFA0300804) and National Science Foundation of China (51502007 and 51672007). A portion of this work was performed at the National High Magnetic Field Laboratory, which is supported by the National Science Foundation Cooperative Agreement No. DMR-1157490 and DMR-1644779 and the State of Florida.

## Author information

These authors contributed equally: Sehun Seo, Heesung Noh

**Affiliations**

*School of Materials Science and Engineering, Gwangju Institute of Science and Technology, Gwangju, 61005, Republic of Korea.*

Sehun Seo, Heesung Noh, Jongmin Lee & Sanghan Lee

*Electron Microscopy Laboratory, and International Center for Quantum Materials, School of Physics, Peking University, Beijing 100871, China*

Ning Li, Ruochen Shi, Mengchao Liu & Peng Gao,

*Applied Superconductivity Center, National High Magnetic Field Laboratory, Florida State University, Tallahassee, Florida 32310, USA*

Jianyi Jiang, Chiara Tarantini & Eric E. Hellstrom,

*Center for Quantum Materials and Superconductivity (CQMS) and Department of Physics, Sungkyunkwan University, Suwon 16419, Republic of Korea*

Soon-Gil Jung & Tuson Park,

*Department of Physics, Kyungpook National University, Daegu, 41566, Republic of Korea*

Myeong Jun Oh & Youn Jung Jo

*Condensed Matter Physics and Materials Science Department, Brookhaven National Laboratory, Upton, New York, 11973, USA*

Genda Gu

**Corresponding authors**

Correspondence to Sanghan Lee

**Conflicts of interest**

There are no conflicts to declare.

# Supplementary Information

**Supplementary information** is available for this paper at NPG Asia Materials'website